\documentclass[aps,prl,twocolumn,showpacs,preprintnumbers,amsmath,amssymb]{revtex4}
\usepackage{graphicx}
\usepackage{epstopdf}
\begin{document}

\title{Practical applications of quantum sensing: a simple method to enhance sensitivity of Nitrogen-Vacancy-based  temperature  sensors}

\author{E. Moreva$^1$, E. Bernardi$^1$, P. Traina$^1$, A. Sosso$^1$, S. Ditalia Tchernij$^{2,3}$, J. Forneris$^{2,3}$, F. Picollo$^{2,3}$, G. Brida$^1$, \v{Z}. Pastuovi\'{c}$^4$,  I. P. Degiovanni$^1$,   P. Olivero$^{2,3,1}$, M. Genovese$^{1,3}$.}
\affiliation{$^1$Istituto Nazionale di Ricerca Metrologica, Strada delle cacce 91, Turin, Italy}
\affiliation{$^2$Physics Department and NIS Centre of Excellence - University of Torino, Torino, Italy}
\affiliation{$^3$Istituto Nazionale di Fisica Nucleare (INFN) Sez. Torino, Torino, Italy}
\affiliation{$^4$Centre for Accelerator Science, Australian Nuclear Science and Technology Organisation, New Illawarra rd., Lucas Heights, NSW 2234, Australia}
\begin{abstract}
Nitrogen-vacancy centers in diamond allow measurement of environment properties such as temperature, magnetic and electric fields at nanoscale level, of utmost relevance for several research fields, ranging from nanotechnologies to bio-sensing.
The working principle is based on the measurement of the resonance frequency shift of a single nitrogen-vacancy center (or an ensemble of them), usually detected by by monitoring the center photoluminescence emission intensity. Albeit several schemes have already been proposed, the search for the simplest and most effective one is of key relevance for real applications.
Here we present a new continuous-wave lock-in based technique able to reach unprecedented  sensitivity in temperature measurement at micro/nanoscale volumes (4.8 mK/Hz$^{1/2}$ in $\mu$m$^3$). Furthermore, the present method has the advantage of being insensitive to the enviromental magnetic noise, that in general introduces a bias in the temperature measurement.
\end{abstract}
\newcommand{\ket}[1]{\mbox{\ensuremath{|#1\rangle}}}
\newcommand{\bra}[1]{\mbox{\ensuremath{\langle#1|}}}
\flushbottom
\maketitle
\thispagestyle{empty}

Reliable techniques for high sensitivity/nanoscale sensing, eventually exploiting the peculiar properties of quantum systems \cite{rev1,rev2}, are of absolute importance for several applications ranging from nanotechnology to biophysics. Among them of particular interest are the innovative approaches to local temperature sensing. These include scanning probe microscopy \cite{yue,vreug}, Raman spectroscopy \cite{raman}, and fluorescence-based measurements using nanoparticles \cite{nanop,nanop2} and organic dyes \cite{dyes}. For example, fluorescent polymers and green fluorescent proteins have recently been used for temperature mapping within a living cell \cite{polymer}. However, many of the existing methods have drawbacks such as poor temporal/spatial resolution, low signal-to-noise ratio, instability in the fluorescence emission and limited operation time. In this perspective, negatively-charged nitrogen vacancy (NV) centers in diamond attracted increasing attention thanks to their unique sensing capabilities \cite{rev1,gop,forn2018}. Sensing techniques based on NV centers in diamond can be implemented at room temperature, they are also the only technique suited for operating in extreme pressure conditions \cite{nat}. Furthermore they have already been applied inside living cells \cite{cells}, a feature that makes them ideal sensors for nanoscale bio-applications. By taking advantage of the biocompatibility \cite{biotox} or anyway of the absence of substantial effects on cell functionality \cite{nos}, neuronal cells can be grown directly on diamond surface \cite{biotox2} or nanodiamonds can be internalized within the cellular body through endocytosis \cite{nos}. Thanks to the small dimensions of nanodiamonds, mapping temperature fluctuations up to sensitivity of 200 mK/Hz$^{1/2}$  and at length scales as small as 200 nm is possible in living cells when being integrated with optical microscopic imaging techniques \cite{kuksco,pt}. Thus, among other sensing applications, NV-based thermometry is attracting increasing interest \cite{acosta,chen,dohe,neumann,wang,toyli,clev,tzeng,andersen,pl}.

Considering the ground triplet state, the degenerated m$_s$ = $\pm$ 1 spin states in the absence of the external magnetic field are separated from the m$_s$ = 0 state by the zero-field splitting (ZFS) parameter, D$_{gs}\sim 2.87$ GHz at room temperature, due to spin-spin interactions in the NV's orbital structures. The  ZFS parameter depends on the lattice spacing, which is strongly influenced by the local temperature. For instance, when the local temperature increases the distance between the localized spins at the NV center increases, lowering the spin-spin interaction and reducing D$_{gs}$. The ZFS parameter non-linearly depends on temperature \cite{chen}, under ambient conditions the temperature dependence is $c_\tau=dD_{gs}/dT \approx$ -74 kHz/K \cite{acosta}. For a sensor containing N color centers, the temperature sensitivity is given by $\eta=\frac{1}{ c_\tau}\frac{\Delta \nu}{C\sqrt{I_0}\sqrt{N}}$, where $\Delta \nu$  and $C$ are respectively the spectral width and the contrast of the spin-resonance dip and $I_0$ is the rate of detected photons. Assuming $\Delta \nu$ of the order of 1 MHz and $C \approx$ 0.3, a single NV can potentially exhibit a sensitivity better than 0.76 K/ Hz$^{1/2}$\cite{dreau}. This values of sensitivity can be improved to 5 mK/ Hz$^{1/2}$ \cite{neumann} using advanced protocols  based on pulsed excitation/control \cite{1,2}, e.g. based on Rabi oscillations and Ramsey sequences \cite{neumann,wang,toyli}, or frequency modulation \cite{tzeng,andersen}. Furthermore, to perform reliable and high-sensitivity temperature measurement the system requires magnetic insulation (for instance from earth magnetic field fluctuations or parasitic magnetization), since the effects of temperature and magnetic field on the resonant frequencies cannot be easily decoupled\cite{andersen,neumann}. This, tipically, increases the encumbrance of the setup.

In this work we present a novel method for nanothermometry based on NV centers in diamonds, paving the way to practical measurement at nanoscale thank to its simplicity of implementation. Our proof-of-principle experiment is performed in bulk diamond, but the method can be easily extented to the nanodiamonds case without significant changes. Specifically, we exploit a sensing scheme with a transverse magnetic field (TF). This results in the suppression of the splitting due to hyperfine interaction, since in this case the expectation value of the spin is small, reducing the coupling between the external magnetic field and the hyperfine field due to the nitrogen nucleus (for a detailed theoretical framework refer to the Supplementary Material). This regime, albeit already reported in literature \cite{dolde2011}, is used here for the first time in temperature sensing, and it can be advantageous in this application on two grounds (see Fig.1): (i) the dips of the optically detected magnetic resonance(ODMR) spectrum have higher contrast and narrower spectral width with respect to common CW ODMR spectra, due to the degenerated hyperfine structure (ii) the presence of a single peak allows the possibility of probing the points of maximum derivative of the ODMR spectrum in frequency modulation scheme. 
It is worth remarking that the degeneration of the hyperfine structure can be clearly observed only if the perpendicular component of the magnetic field, $B_\perp$, is strong enough to separate a a pair of symmetric dips on a frequency difference which is larger than the hyperfine splitting\footnote{This corresponds to $B_\perp=\frac{h}{g\mu_B}\sqrt{\Delta A_{\parallel} D_{gs}}$, where $A_{\parallel}=2.16$ MHz is the parallel hyperfine interaction with the $^{14}$N nuclear spin, $g$ is Land\'{e} factor and $\mu_B$ is Bohr Magneton. Thereby the boundary condition is $B_\perp>2.8$ mT. In our case the contribution from not axial strain is negligible}.
A common  technique used to increase the sensitivity of ODMR measurements simultaneously probes all three hyperfine peaks related to one spin orientation (simultaneous hyperfine driving, SHfD) instead of probing just one of them. Typically this methods increase the sensitivity of a  factor of $\sim$ 2.

\begin{figure}[!h]
\centering
\includegraphics[scale=0.35]{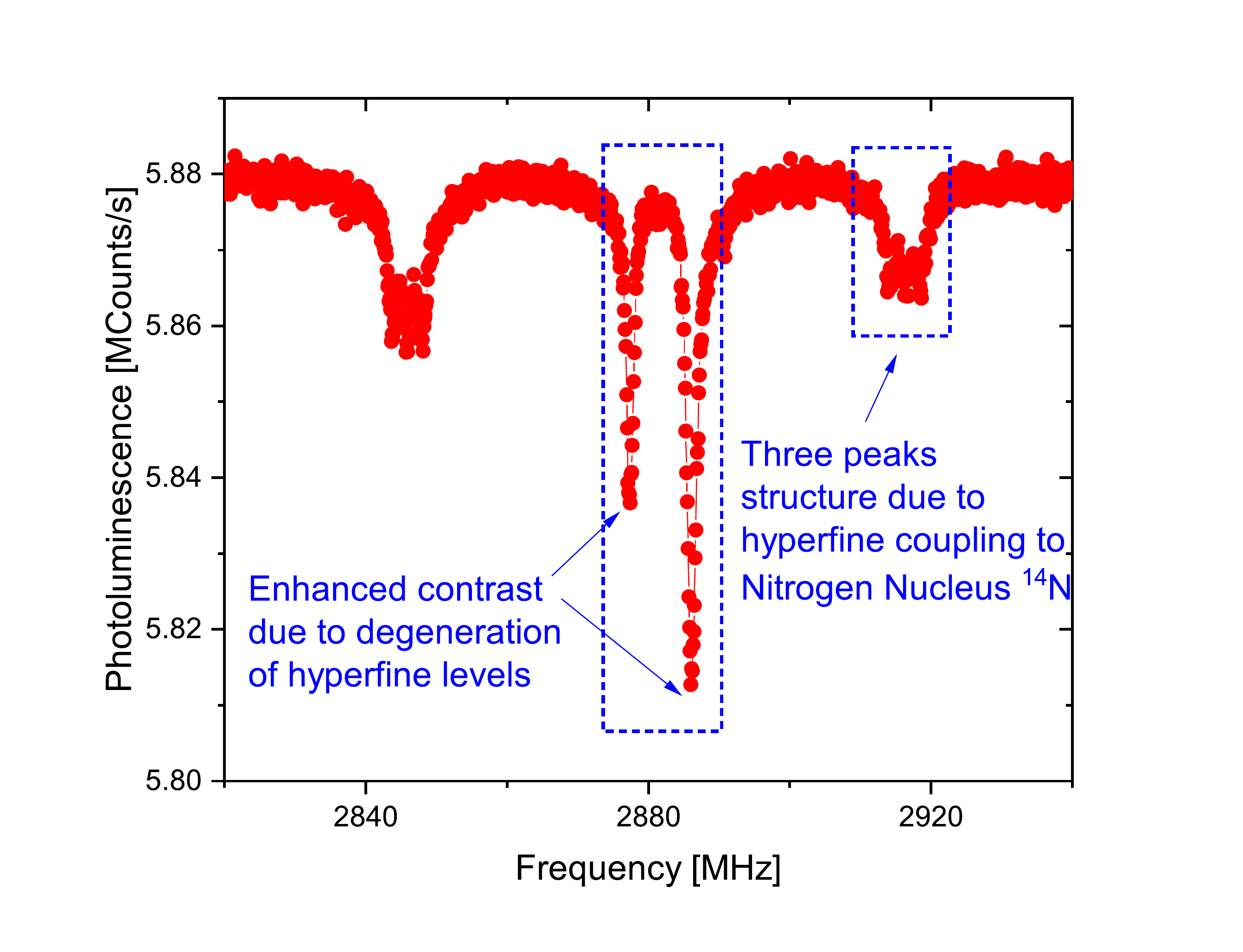}
\caption{Optically Detected Magnetic Resonance spectra: orientating the magnetic field $\textbf{B}$ orthogonally to NV axis suppresses hyperfine contribution resulting in increased contrast and reduced linewidth of the resonant peak and hence in enhanced sensitivity.}
\label{fig:odmr}
\end{figure}

\begin{figure}[!h]
\centering
\includegraphics[scale=0.3]{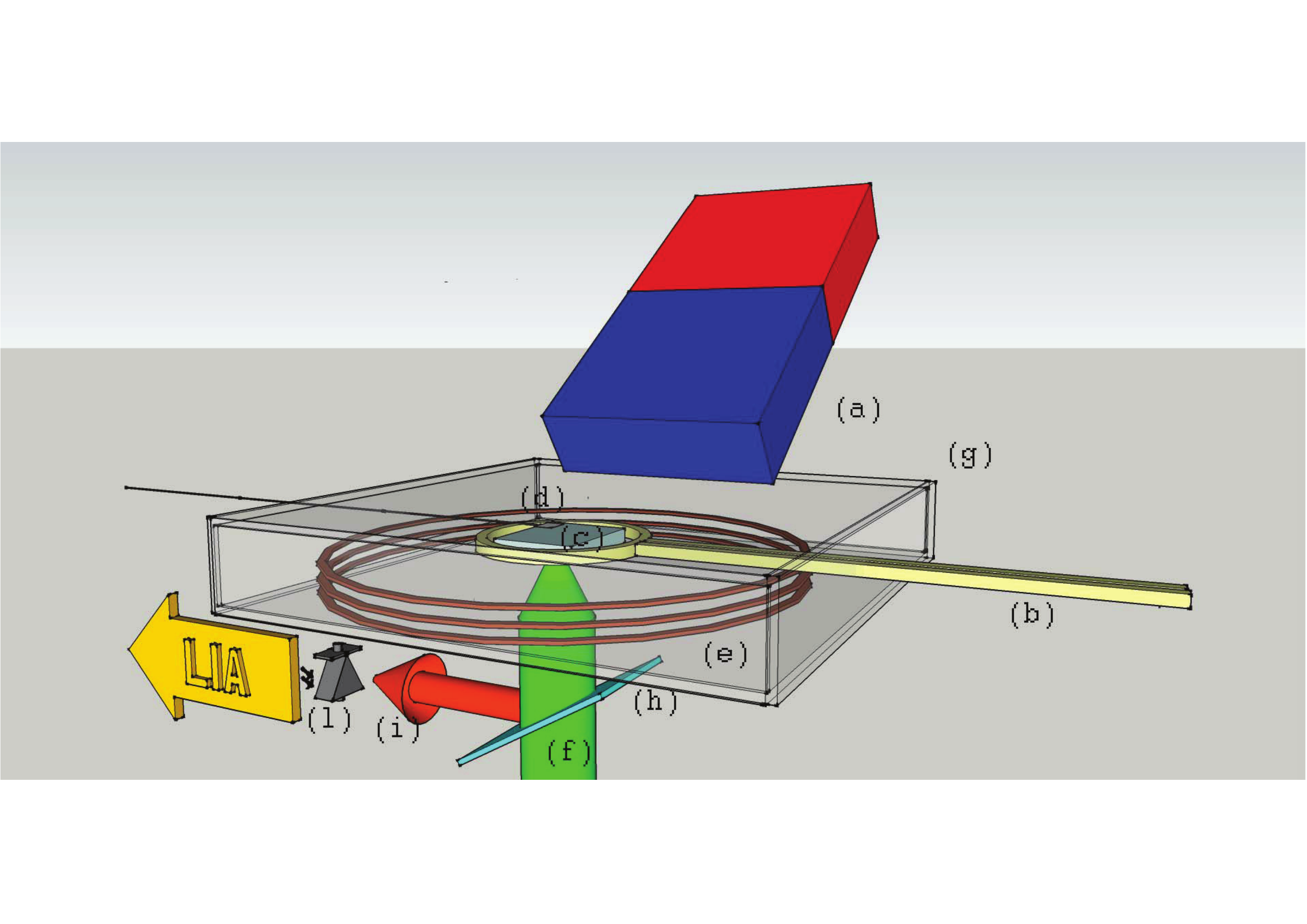}
\caption{Scheme of the experimental setup. (a) Permanent magnet (b) Microwave planar ring antenna (c) CVD diamond sample (d) Thermocouple (e) Coil (f) Green Excitation laser (g) Temperature-controlled chamber (i)Red Photoluminescence (l) Photodetector }
\label{fig:setup}
\end{figure}

Our setup (Fig.2) is based on of an inverted microscope (Olympus IX73), adapted to confocal measurements with single-photon sensitivity. The sensor consisted of a CVD diamond sample by Element Six $3\cdot3\cdot0.3$ mm$^{3}$ substrate, having $<1$ ppm and $<0.05$ ppm concentrations of substitutional nitrogen and boron, respectively. The NV sensing volume was fabricated upon the implantation of 10 keV $^{14}$N$^{+}$ ions with a fluence of 10$^{14}$ ions $\cdot$ cm$^{-2}$ followed by a thermal annealing at 950 °C for 2 hours. The process resulted in the formation of a $\sim 10$ nm thick layer of NV centers with a concentration of rougly $n_{NV} \sim 3 \cdot 10^{19}$ cm$^{-3}$. The diamond sample was mounted on a microwave planar ring antenna, specifically designed for ODMR measurements at frequencies within a 400 MHz range centered around the 2.87 GHz spin resonance\cite{mw}, in the  temperature-controlled chamber (293.15-318.15 K, $\pm$ 0.1 K) of the the microscope.  As a reference, temperature measurement with a precision of $5\cdot10^{-3}$ K was also carried on, exploiting a thermocouple fixed on the top of the diamond by means of thermo-conductive glue.
Excitation light (80 mW optical power) at 532 nm was obtained by the second harmonic of a Nd:YAG laser with high power stability (Coherent Prometheus 100NE) and  was focused close to the lower side of the diamond sample through an air objective (Olympus UPLANFL) with Numerical Aperture NA=0.67. The photoluminescence (PL), spectrally filtered with notch and long-pass filter at 650 nm, was collected and detected with two different acquisition systems. A 4\% fraction of the total PL intensity was sent to a single photon detector (SPD). The signal from the SPD was used for the ODMR spectrum acquisition. The remaining 96\% fraction  of the emitted PL intensity was collected by NA=0.25 objective (Olympus 10$\times$) and imaged onto a bias photodetector (Thorlabs DET 10A2). An external magnetic field was applied to the diamond sample using a permanent magnet fixed on a 
translation stage allowing micrometric movement along the three spatial axes. The minimum magnet-sample distance achievable with our apparatus was 4 cm because of the thickness of the chamber. In our case, the magnetic field  was approximately 6 mT. In addition a coil was used for magnetic field modulation.
The microwave control was obtained by a commercial microwave (MW) generator (Keysight N5172B) whose central frequency was internally modulated at 
$f_{mod} = 1009$ Hz. The  output MW signal was amplified and then sent to the microwave antenna.
The photodiode  was  connected to the low-noise current input of a lock-in amplifier(LIA) (Stanford SR860) which provided phase-sensitive demodulation of the fluorescent signal.
In order to operate in SHfD regime, the MW was  mixed via a double balanced mixer with a $\sim$ 2.16 MHz sinewave  to create two simultaneous driving modulated frequencies near the selected resonance transition.

\begin{figure}[!h]
\centering
\includegraphics[scale=0.37]{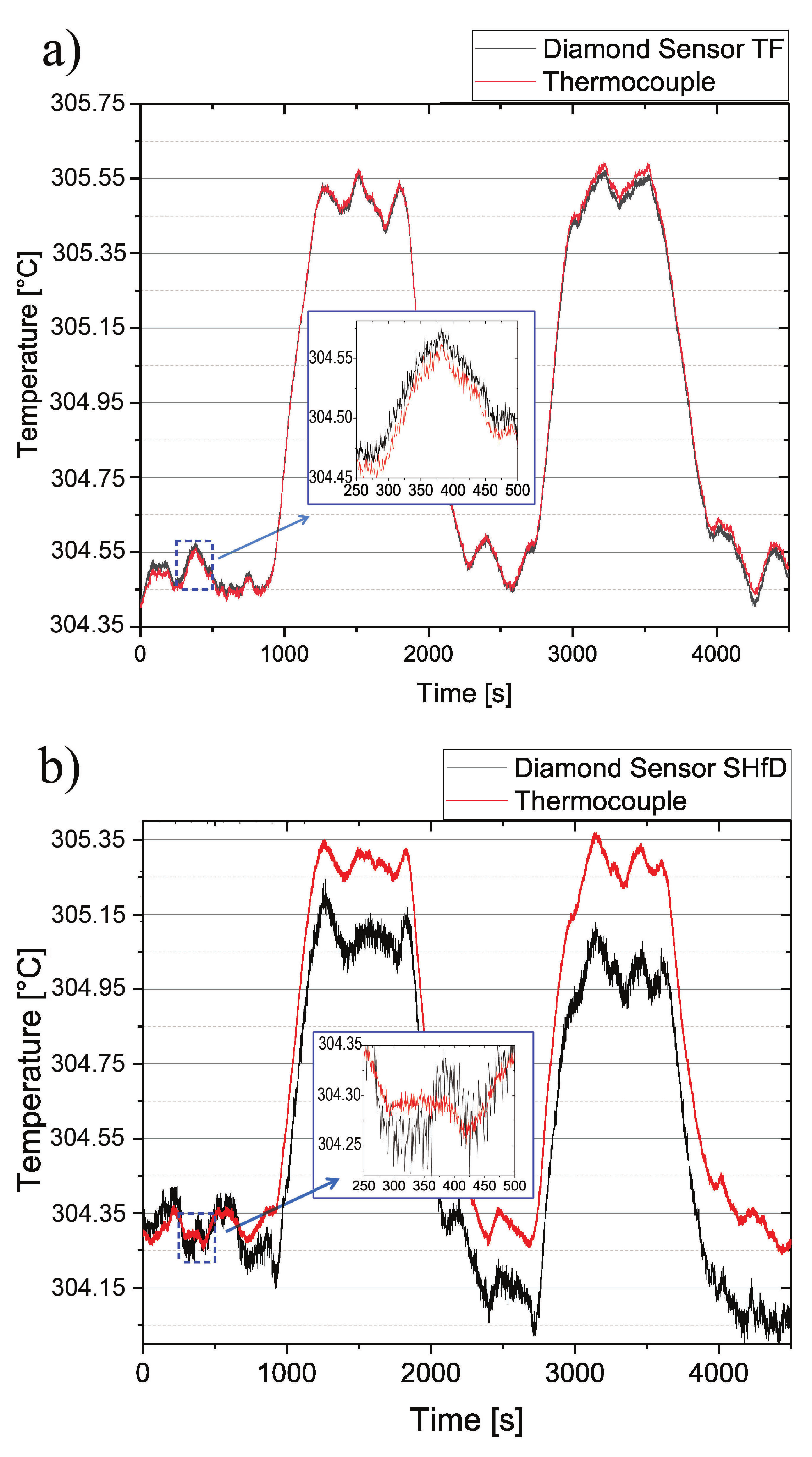}
\caption{Response of the NV sensor (black line) in the proposed regime compared to the readout of a standard thermocouple (red line) to repeated thermal cycles, showing almost perfect agreement of the measurement values. Inset a) refers to the transverse magnetic field (TF) regime; inset b) refers to the simultaneous hyperfine driving(SHfD). A better agreement between the NV sensor and the thermocouple and a reduced uncertainty are observed in the former case. The lack of agreement between the two traces in the latter case is attributed to environmental  field fluctuations. The period of the repetition is 1800 s and the temperature variation is 1 K.}
\label{fig:step}
\end{figure}

\begin{figure}[!h]
\centering
\includegraphics[scale=0.32]{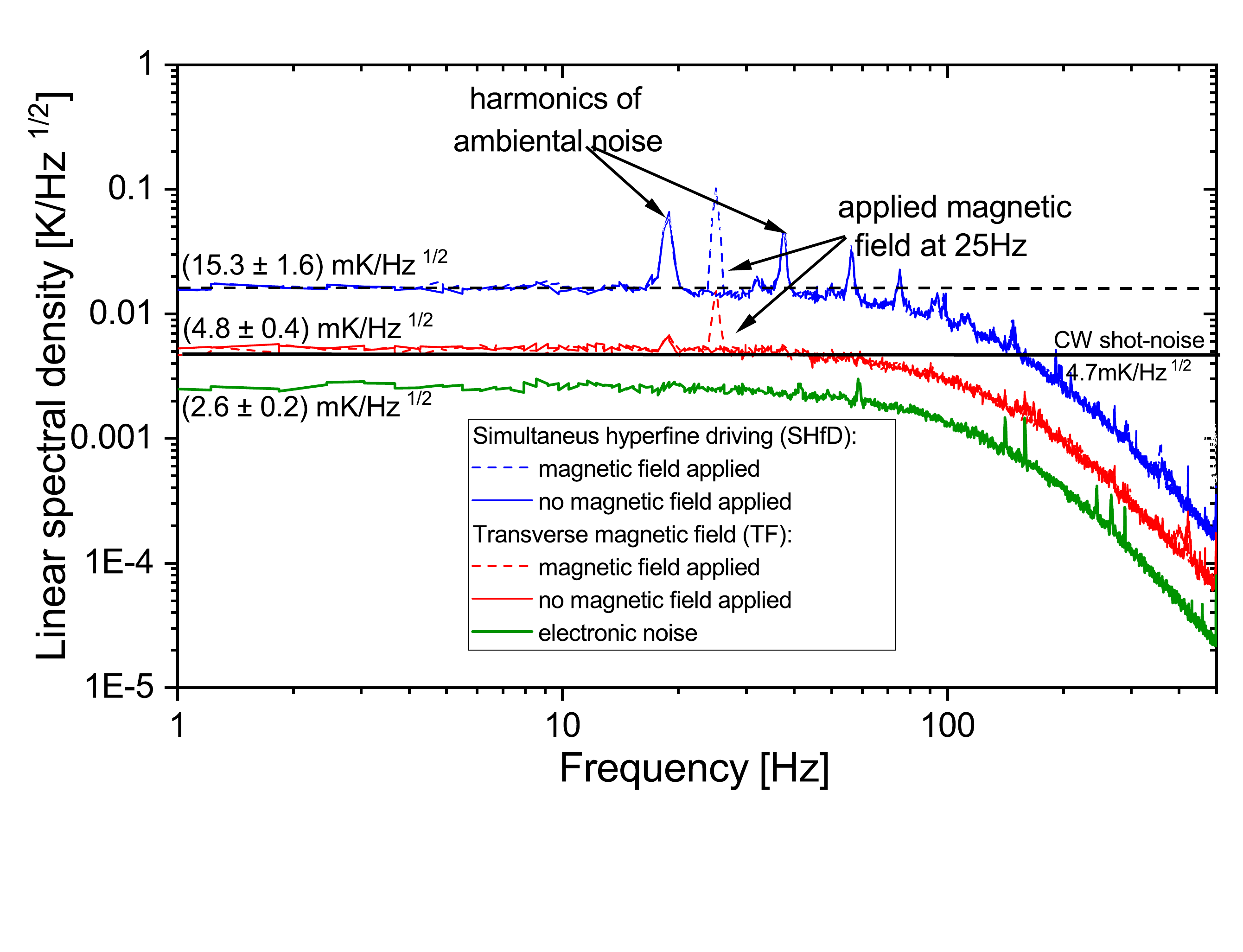}
\caption{Comparison between the linear spectral densities of the readout of the NV sensor in the transverse magnetic field (TF) regime (red line) and in the simultaneous hyperfine driving(SHfD) case (blue line). The dashed/continuous lines correspond, respectively, to the application/disabling of an oscillating (25 Hz) magnetic field to the sample. It is apparent from the continuous lines that an improved sensitivity (lower than 5 mK/Hz$^{1/2}$) is achieved with our method. From the dashed lines it is demonstrated that with our technique the peak  due to the 25 Hz magnetic field is one order of magnitude less prominent in our regime, proving the reduced sensitivity to magnetic noise.}
\label{fig:floor}
\end{figure}

The measurement required a preliminary calibration to determine the parameter D$_{gs}$ dependence on temperature. This calibration was carried by determining the MW frequency for which LIA signal is zero for different temperatures measured with the thermocouple. From this calibration a value of $c_\tau=-74.2$ kHz/K is obtained. The actual temperature measurement was carried by fixing the MW frequency and recording the variation of LIA signal in time (averaging time 1 s, lock-in time constant $\tau=$ 10 ms). This is related to the variation in temperature through the slope of the LIA spectrum and the parameter $c_\tau$. With this independent calibration of the temperature dependence on D$_{gs}$, we carried the temperature measurement by exploiting both our technique and the one based on SHfD, as respectively reported in fig.s 3(a) and 3(b).   
Fig. \ref{fig:step}(a) shows the response of the NV sensor (black line) to repeated thermal cycles (1800 s period, 1 K temperature variation) compared to the readout of a standard thermocouple (red line). The picture shows the excellent agreement between these two measurements performed with independent devices. We point out that the uncertainty of the SHfD method in Fig. \ref{fig:step}(b) is about 3 times larger than for the transverse bias regime. Furthermore, we note that in the latter case there is no agreement between the two traces at increasing times. This is due to periodic fluctuations of environmental magnetic field. Instead, in our technique the contribution of the magnetic field appears only at the second order in the Hamiltonian, thus protecting the measurements from fluctuations of environmental magnetic field.
In order to characterize the sensitivity of our temperature measurement method, time traces of the LIA output signal were recorded and a root-mean square amplitude spectral density was calculated using the Hanning window function. Noise contributions arising from laser, microwave excitation and electronic noise of the lock-in amplifier are accounted for. In addition, CW shot-noise limit was calculated, as described in the Supplementary Material.
The MW frequency was tuned to the center of the resonance, where the lock-in signal crossed zero and thus provided the largest temperature response. The sinewave modulation at 1009 Hz and modulation depth 0.6 MHz were used for all recordings.
The DC plateau of the curves expresses the experimental sensitivity and in our case, by fitting the curve, a sensitivity value  as low (4.8 $\pm$ 0.4) mK/Hz$^{1/2}$ is demonstrated.
In Fig. \ref{fig:floor}, showing the linear spectral density of the sensed signal (acquisition time 10 min, lock-in time constant $\tau=$ 1 ms), we note that the sensitivity obtained with our method (red  curves) with respect to the one achieved by SHfD (blue curves) is enhanced by a $\sim 3$ factor, coherently with the result in Fig. 3. 
 The electronic noise floor is about 2 mK/Hz$^{1/2}$. The other major noise contribution comes from laser fluctuations, that results near to CW shot-noise limit: thus the "physical" limit to our innovative method is significantly lower. To further demonstrate the resilience of our technique to external magnetic field, we injected an additional oscillating (25 Hz) magnetic field. The effect of this magnetic field is significantly different for the two techniques, as it can be seen in Fig. 4, where the peak in the spectrum corresponding to this magnetic signal is one order of magnitude less prominent with our technique as compared to the SHfD method \footnote{The oscillating magnetic field ($\approx 100$ nT) is applied along (100) direction, resulting in projection of equal magnitude along every NV-axis orientation}. This is consistent with the fact that the main harmonics present in SHfD spectrum substantially disappear, demonstrating that our technique is less sensitive to magnetic noise. In our proof-of-principle measurements we used a bulk diamond, that allows the measurement of the average temperature through the whole sample due to the high thermal conductivity of this material ($\lambda=2500$ W$\cdot$m$^{-1}\cdot$K$^{-1}$). This was necessary to compare our results with the reading of the termocouple. Practical nanoscale sensing can be obtained with this technique by using nanodiamond sensors.

In conclusion, in this letter we presented a new technique for temperature sensing with NV centers in diamond, allowing reaching a high sensitivity without the need of a complex experimental setup, such as for instance the implementation of articulated pulse sequences, while at the same time protecting the measurement from the effect of enviromental magnetic  noise (which represents one of the most significant limit of present techniques). 
Indeed, our technique only requires the alignment of the bias magnetic field along the orthogonal direction of the NV axis. In particular we have obtained $<$ 5 mK/Hz$^{1/2}$ noise floor  which represents a value comparable with the best results achieved up to now \cite{andersen}( 430 $\mu$ K/Hz$^{1/2}$) when the sensitivity is normalized to the sensing volume ($\sim 1\mu$m$^3$ in our case, $\sim 5 \cdot 10^3 \, \mu$m$^3$ in \cite{andersen} ).  On the other hand, techniques based on ZPL shift \cite{ZPL1,ZPL2,ZPL3,ZPL4,ZPL5,ZPL6}, albeit simpler, are not competitive in terms of sensititivity (usually about 500 mK/Hz$^{1/2}$), eventually reduced to  13 mK/Hz$^{1/2}$ but at the price of a more complex measurement protocol\cite{ZPL3}, i.e.  a multiparametric analysis of photolumiscence, and acquisition time.

In a short time perspective this technique can be extended to nanoscale or intracellular sensing by using nanodiamonds instead of a bulk crystal. A further straightforward extension of the method would be based on the use of the degenerated resonances  as a close-loop control for the other, non-degenerate, resonances allowing for simultaneous, independent monitoring of both the magnetic field and the temperature. Finally, this method thank to its simplicity and high sensitivity could boost the development in the field of quantum-assisted temperature sensing and it has foreseeable applications in high-sensitivity nanoscale thermometry and, thanks to the biocompatibility of diamond, biosensing.

\section*{Acknowledgements}

This work has received funding from the European Union's PATHOS EU H2020 FET-OPEN grant no. 828946 and Horizon 2020 and the EMPIR Participating States in the context of the projects EMPIR-17FUN06 ''SIQUST'' and 17FUN01 BeCOMe, from the project ''Piemonte Quantum Enabling Technologies'' (PiQuET), funded by the Piemonte Region within the "Infra-P" scheme (POR-FESR 2014-2020 program of the European Union), from "Departments of Excellence" Project (L. 232/2016), funded by MIUR, from Coordinated Research Project "F11020" of the International Atomic Energy Agency (IAEA), from the Australian Government through the NCRIS programme.

The authors acknoledge the technical support received from ÏElio Bertacco, Vanna Pugliese, Simone D'Agostino, Fabio Bertiglia, Giuseppina Lo Pardo.

\section{Supplementary Material}

\subsection{Effect of orthogonal magnetic field}

\begin{figure*}
\centering
\includegraphics[scale=0.15]{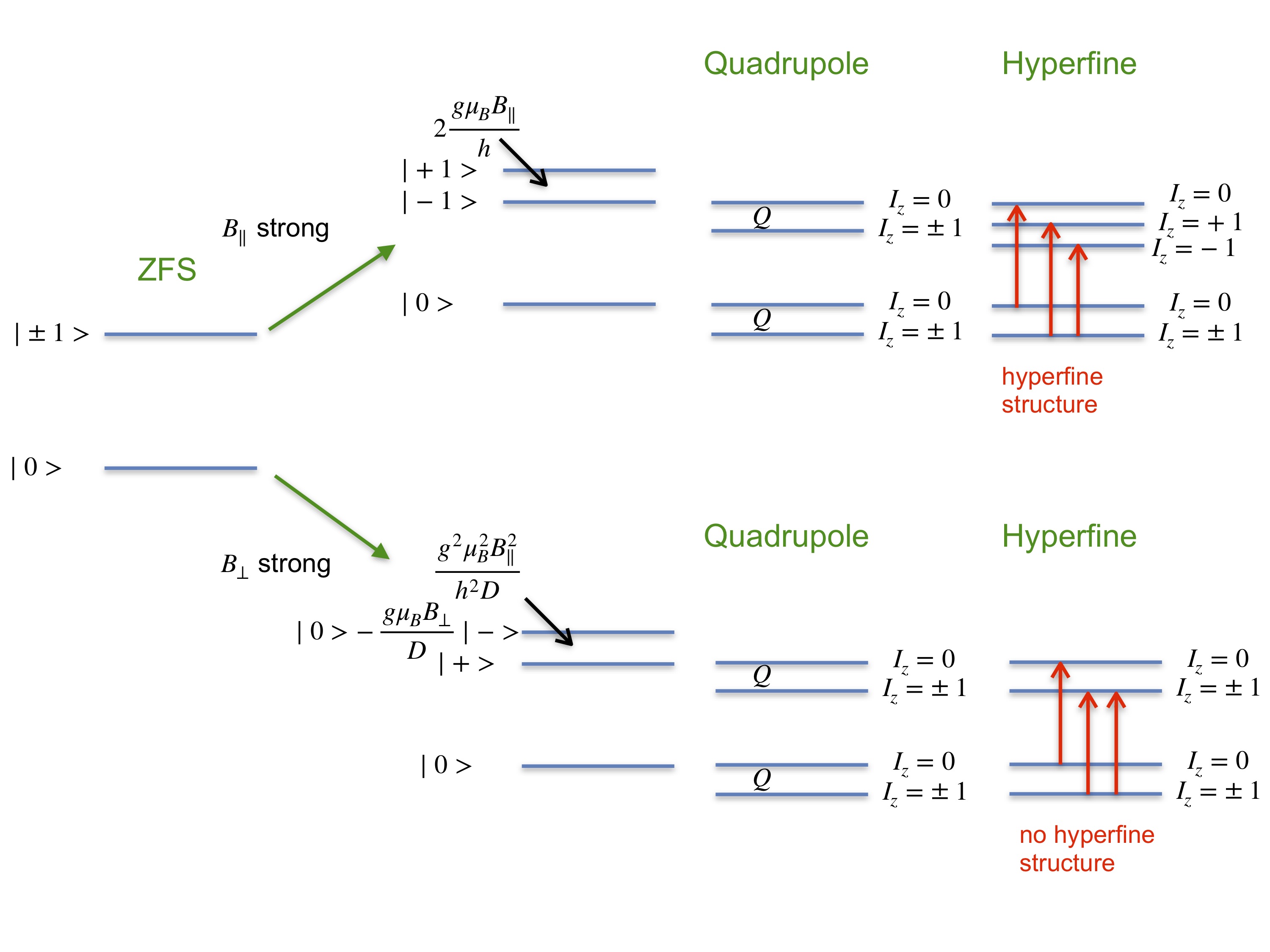}
\caption{Energy levels splitting for axial and tranverse magnetic field}
\label{fig:Hyperfine}
\end{figure*}

Considering the unperturbed Hamiltonian for a single NV center:

\begin{equation}
\label{eq:Ham_unp}
H_0 = DS_z^2
\end{equation}

with the unperturbed eigenvectors and eigenvalues 

\begin{align}
\label{eq:eigenvalues_unpert}
\begin{split}
\ket{0} = \ket{S_z=0} \,,\, 0\\ 
\ket{+} = \frac{1}{\sqrt{2}}(\ket{S_z=+1} + \ket{S_z=-1}) \,,\, D \\
\ket{-} = \frac{1}{\sqrt{2}}(\ket{S_z=+1} - \ket{S_z=-1}) \,,\, D\\
\end{split}
\end{align}

considering the effect of an orthogonal magnetic field, that we consider aligned along $x$ without any loss of generality assuming a isotrope system in the plane xy

\begin{equation}
\label{eq:Ham_pert}
H_{\bot}=\frac{g\mu_B}{h}B_xS_x\\
\end{equation}

The matrix of the complete Hamiltonian in the basis of unperturbed eigenvalues, see Eq. \ref{eq:eigenvalues_unpert}, will be:

\begin{equation}
\label{eq:Ham_matrix_complete}
\begin{pmatrix}
0 & 0 & \frac{g\mu_B}{h}B_{x}\\
0 & D & 0\\
\frac{g\mu_B}{h}B_{x} & 0 & D\\
\end{pmatrix}
\end{equation}

If  $ \frac{g\mu_BB_{\bot}}{D}\ll1 $ we can apply the second-order perturbation theory for degenerate states.
The resulting perturbed eigenstates and eigenvectors are respectively:

\begin{align}
\label{eq:eigenvalue_pert}
\begin{split}
E_1 = -\frac{g^2\mu_B^2B_{\bot}^2}{h^2D}\\ 
E_2= D\\
E_3= D + \frac{g^2\mu_B^2B_{\bot}^2}{h^2D}\\
\end{split}
\end{align}

and:

\begin{align}
\label{eq:eigenvector_pert}
\begin{split}
\ket{1} = \frac{1}{\sqrt{1+\frac{g^2\mu_B^2B_{\bot}^2}{h^2D^2}}}(\ket{0} - \frac{g\mu_BB_{\bot}}{D}\ket{+})\\ 
\ket{2} = \ket{-}\\
\ket{3} = \frac{1}{\sqrt{1+\frac{g^2\mu_B^2B_{\bot}^2}{h^2D^2}}}(\frac{g\mu_BB_{\bot}}{D}\ket{0} + \ket{+})\\
\end{split}
\end{align}

the state $\ket{-}$ remains unperturbed, the states $\ket{S_z=0}$ and $\ket{+}$ are mixed and repel each others.  

From Eq. \eqref{eq:eigenvalue_pert} follows that the transitions $\ket{1}\rightarrow\ket{2}$ and  $\ket{1}\rightarrow\ket{3}$ are both shifted to higher values:
\begin{align}
\label{eq:trans_shifting}
\begin{split}
\Delta_{\ket{1}\rightarrow\ket{2}} = D + \frac{g^2\mu_B^2B_{\bot}^2}{D} = E_2 - E_1 \\
\Delta_{\ket{1}\rightarrow\ket{3}} = D + 2\frac{g^2\mu_B^2B_{\bot}^2}{D} = E_3 - E_1\\
\end{split}
\end{align}

It is worth calculate the expectation value of the spin in different direction  on the perturbed states \eqref{eq:eigenvector_pert}

For the spin along $x$ we have:

\begin{align}
\label{eq:eigenstates}
\begin{split}
\bra{1}S_x\ket{1} =-2\frac{g\mu_BB_{\bot}}{D}\frac{1}{1+\frac{g^2\mu_B^2B_{\bot}^2}{D^2}} \\
\bra{2}S_x\ket{2}=\bra{-}S_x\ket{-}=0 \\
\bra{3}S_x\ket{3} = +2\frac{g\mu_BB_{\bot}}{D}\frac{1}{1+\frac{g^2\mu_B^2B_{\bot}^2}{D^2}} \\
\end{split}   
\end{align}
 
For the spin along $y$ we have:

\begin{equation}
\bra{1}S_y\ket{1}=\bra{2}S_y\ket{2}=\bra{3}S_y\ket{3}=0    
\end{equation}

For the spin along $z$ we have:

\begin{equation}
\bra{1}S_z\ket{1}=\bra{2}S_z\ket{2}=\bra{3}S_z\ket{3}=0    
\end{equation}

From these equations results that expectation value of the spin is null along the directions orthogonal to applied field, and small along the direction parallel to the applied field.
This imply that also the magnetic moment, and hence the coupling with external magnetic field and hyperfine field due to the nitrogen nucleus is small.

The $^{14}$N nuclear contribution  can be expressed as:

\begin{equation}
\label{eq:ham_nucl}
\mathcal{H}_{nucl}=QI_z^2  + S_zA_{\parallel}I_z + S_xA_{\bot}I_x + S_yA_{\bot}I_y
\end{equation}

where $Q$ is quadrupole coupling and $A_{\parallel}$,$A_{\bot}$ the parallel and normal hyperfine couplings. 

We will consider the case of $B_{\bot}$ along $x$ and calculate the contribution of the nuclear terms to perturbed eigenstates \eqref{eq:eigenvector_pert} as the expectation value of \eqref{eq:ham_nucl} on these states.
It can be shown that the quadrupular contribution acts only on nuclear degrees of freedom creating a gap between states with $I_z=0$ and states with $I_z \pm 1$ 

Considering the hyperfine term for the $\ket{I_z=+1}\ket{3}$ state

\begin{multline}
\label{eq:expec_hyperf}
 \bra{3}\bra{I_z=+1}\mathcal{H}_{hyper}\ket{I_z=+1}\ket{3}=\\
\bra{3}\bra{I_z=+1}S_zA_{\parallel}I_z + S_xA_{\bot}I_x + S_yA_{\bot}I_y\ket{I_z=+1}\ket{3}=\\
\bra{3}\bra{I_z=+1}S_zA_{\parallel}I_z\ket{I_z=+1}\ket{3}+\bra{3}\bra{I_z=+1}S_xA_{\bot}I_z\ket{I_z=+1}\ket{3}+\bra{3}\bra{I_z=+1}S_yA_{\bot}I_z\ket{I_z=+1}\ket{3}=\\
A_{\parallel}\bra{3}S_z\ket{3}\bra{I_z=+1}I_z\ket{I_z=+1}+A_{\bot}\bra{3}S_x\ket{3}\bra{I_z=+1}I_x\ket{I_z=+1}+\\
+A_{\bot}\bra{3}S_y\ket{3}\bra{I_z=+1}I_y\ket{I_z=+1}=0
\end{multline}

because the only not null term for electronic degrees of freedom is  $\bra{3}S_x\ket{3}$, but in this case is null the nuclear term $\bra{I_z=+1}I_x\ket{I_z=+1}$. 
Similarly:

\begin{align}
\begin{split}
\bra{3}\bra{I_z=0}\mathcal{H}_{hyper}\ket{I_z=0}\ket{3}=0\\
\bra{3}\bra{I_z=+1}\mathcal{H}_{hyper}\ket{I_z=+1}\ket{3}=0\\
\end{split}
\end{align}

Along the same lines it can be shown that:

\begin{align}
\begin{split}
\bra{2}\bra{I_z=0}\mathcal{H}_{hyper}\ket{I_z=0}\ket{2}=0\\
\bra{2}\bra{I_z=+1}\mathcal{H}_{hyper}\ket{I_z=+1}\ket{2}=0\\
\bra{2}\bra{I_z=-1}\mathcal{H}_{hyper}\ket{I_z=-1}\ket{2}=0\\
\end{split}
\end{align}

and

\begin{align}
\begin{split}
\bra{1}\bra{I_z=0}\mathcal{H}_{hyper}\ket{I_z=0}\ket{1}=0\\
\bra{1}\bra{I_z=+1}\mathcal{H}_{hyper}\ket{I_z=+1}\ket{1}=0\\
\bra{1}\bra{I_z=-1}\mathcal{H}_{hyper}\ket{I_z=-1}\ket{1}=0\\
\end{split}
\end{align}

The main point is that the quadrupole interaction align nuclear spin along NV axis but perturbed electronic states have a small electronic spin different from zero only in a direction orthogonal to NV axis. Thus, for a magnetic field aligned orthogonal to the NV axis the hyperfine interaction has no effect on the level structure and hence no hyperfine structure is present, see Fig. \ref{fig:Hyperfine}.

\subsection{Shot noise limit calculation}


The CW shot-noise limit is:

\begin{equation}\label{eq:sensitivity}
\eta_{CW}=\frac{1}{c_\tau}\frac{\sqrt{I_0}}{\max\left(\frac{\partial I_0}{\partial\nu}\right)}
\end{equation}

where $c_{\tau}=74.2$ kHz/K.

We can estimate it from data in fig. 1 of the main text considering the equivalent expression

\begin{equation}\label{eq:sensitivity_bis}
\eta_{CW}=K\frac{1}{c_\tau}\frac{\Delta\nu}{\sqrt{I_0}C}=4.7 \,\, \text{mK/Hz}^{1/2}
\end{equation}

where $\Delta\nu=1.1$ MHz is the linewidth, $C=0.64\%$ the contrast for the left hand side ODMR peak related  to strong transverse magnetic field, see Fig.1 in the main text, and $K=0.31$ is specific of our linewidth. $I_0=3.03*10^{10}$ s$^{-1}$ is estimate from the optical power incident  to the photodiode $W=8.5$ nW, considering a photon energy $E_{ph}=2.84\cdot10^{-19}$ J.  It is worth noticing that using lock-in detection the points of maximum derivative are probed on both sides of the peak. This implies that the signal increase by a factor 2 and the noise by a factor $\sqrt{2}$ leading to a shot noise limit lowered by a factor $\sqrt{2}$.

\end{document}